\documentclass[12pt]{article}

\usepackage{graphics,graphpap}

\renewcommand{\theequation}{\arabic{equation}}
\newcommand{\bea}{\begin{eqnarray}}
\newcommand{\eea}{\end{eqnarray}}
\newcommand{\beq}{\begin{equation}}
\newcommand{\eeq}{\end{equation}}

\def\msbar{\ifmmode{\overline{\rm MS}} \else{$\overline{\rm MS}$} \fi}
\def\drbar{\ifmmode{\overline{\rm DR}} \else{$\overline{\rm DR}$} \fi}
\def\sf{\ifmmode{\tilde{f}} \else{$\tilde{f}$} \fi}
\def\st{\ifmmode{\tilde{t}} \else{$\tilde{t}$} \fi}
\def\sb{\ifmmode{\tilde{b}} \else{$\tilde{b}$} \fi}
\def\sq{\ifmmode{\tilde{q}} \else{$\tilde{q}$} \fi}
\def\sg{\ifmmode{\tilde{g}} \else{$\tilde{g}$} \fi}
\def\bbar{\ifmmode{\bar{b}} \else{$\bar{b}$} \fi}
\def\tbar{\ifmmode{\bar{t}} \else{$\bar{t}$} \fi}
\def\qbar{\ifmmode{\bar{q}} \else{$\bar{q}$} \fi}
\def\ksla{{k \hspace{-2mm} \slash}}

\newcommand\bvec{\left( \begin{array}{c}}
\newcommand\evec{\end{array}\right)}

\newcommand\bmat{\left( \begin{array}{cc}}
\newcommand\emat{\end{array}\right)}

\newcommand\ch{{\tilde{\chi}}}
\renewcommand\d{\delta}

\def\ksla{{k \hspace{-2.2mm} \slash}}

\def\psla{{p \hspace{-1.8mm} \slash}}

\newcommand{\tb}{\tan\beta}

\newcommand\cch{{\tilde\chi^0}}

\newcommand\tw{{\tan{\theta_W}}}

\newcommand\forpi{{\frac1{\left(4\pi\right)^2}}}

\newcommand{\IL}{{I_f^{3L}}}

\newcommand\ca{{\cos{\alpha}}}
\newcommand\sa{{\sin{\alpha}}}
\newcommand\cas{{\cos^2{\alpha}}}
\newcommand\sas{{\sin^2{\alpha}}}
\newcommand\cbe{{\cos{\beta}}}
\newcommand\sbe{{\sin{\beta}}}

\newcommand\cbs{{\cos^2{\beta}}}
\newcommand\sbs{{\sin^2{\beta}}}

\renewcommand\d{\delta}

\newcommand{\M}{{\cal M}}

\def\su{\ifmmode{\tilde{u}} \else{$\tilde{u}$} \fi}
\def\sd{\ifmmode{\tilde{d}} \else{$\tilde{d}$} \fi}

\newcommand{\al}{\alpha}
\newcommand{\be}{\beta}
\newcommand{\si}{{\mbox{s}}}
\newcommand{\co}{{\mbox{c}}}

\begin{document}
%------------------------------------------------------------------------

\pagestyle{empty} \vspace*{-1cm}
\begin{flushright}
%  November 22, 2001\\
  HEPHY-PUB 749/01 \\
  TU-639 \\
  hep-ph/0111303
\end{flushright}

\vspace*{1.4cm}

\begin{center}
\begin{Large} \bf
One-loop corrections to\\[2mm]
 neutral Higgs boson decays into neutralinos
\end{Large}

\vspace{10mm}

{\large H. Eberl$^a$, M. Kincel$^{a,\,b}$, W. Majerotto$^a$,
 Y.~Yamada$^c$}

\vspace{6mm}
\begin{tabular}{l}
 $^a${\it Institut f\"ur Hochenergiephysik der \"Osterreichischen
 Akademie der Wissenschaften,}\\
 \hphantom{$^a$}{\it A--1050 Vienna, Austria}\\
 $^b${\it Department of Theoretical Physics FMFI UK, Comenius
 University,  SK-84248}\\
  \hphantom{$^b$}{\it Bratislava, Slovakia }\\
 $^c${\it Department of Physics, Tohoku University,
Sendai 980--8578, Japan}
\end{tabular}

\vspace{20mm}

\begin{abstract}
We present the one-loop corrected decay widths for the decays of
the neutral Higgs bosons $h^0$, $H^0$ and $A^0$ into a neutralino
pair $\ch^0_m\, \ch^0_n$ $(m, n = 1,\ldots,4)$ and to the decay
$\ch^0_m \to (h^0,\,H^0,\,A^0)\,+\,\ch^0_n$.
The corrections contain the one-loop contributions of all fermions
and sfermions. All parameters are taken on-shell. This requires a
proper treatment of the neutralino mass and mixing matrix. The
dependence on the SUSY parameters is discussed. The corrections
can be large in certain regions of the parameter space.
\end{abstract}
\end{center}

\vfill

\newpage
\pagestyle{plain} \setcounter{page}{2}

\section{Introduction}
The Minimal Supersymmetric Standard Model (MSSM) \cite{mssm} is
considered the most attractive extension of the Standard Model. The
MSSM requires the existence of two isodoublets of scalar Higgs fields,
implying three neutral Higgs bosons, two CP-even bosons ($h^0$, $H^0$),
one CP-odd ($A^0$), and two charged Higgs bosons ($H^{\pm}$).
Searching for these Higgs bosons is one of the main goals of all future
colliders as the Tevatron, LHC, and an $e^+ e^-$ linear collider. The
search strategies very much depend on the way these Higgs bosons decay.
It is therefore mandatory to have a clear picture of the decay modes.
Thus it is necessary to calculate the widths and branching
ratios of the various decays as precisely as possible.

The lightest Higgs boson $h^0$ with a mass of at most $140$ GeV
will decay mainly into $b\bar{b}$ and to a lesser extent into
${\tau}^+{\tau}^-$. It is, however, possible that it also decays as
\begin{equation}
    h^0\to \ch^0_1  + \ch^0_1 \, ,
    \label{eq:h0ch1ch1}
\end{equation}
where $\ch^0_1$ is the lightest neutralino. In the case of $R$-parity
conservation, this decay is invisible, and its appearance would reduce
the branching ratios of the other decay modes. The heavier neutral
Higgs bosons $H^0$ and $A^0$ may decay into a pair of neutralinos
\begin{equation}
  (H^0, A^0) \to \ch^0_m  + \ch^0_n \, ,
  \label{eq:Hkchmchn}
\end{equation}
with $(m,n=1,\ldots,4)$. At tree level, the decays occur by
higgsino-gaugino-Higgs boson couplings~\cite{gunion}, and are
therefore sensitive to the components of neutralinos. The decays
(\ref{eq:h0ch1ch1}) and (\ref{eq:Hkchmchn}) as well as those of
$H^{\pm}\to \tilde\chi^{\pm}_{i}\ch^0_m,\;(i=1,2)$ have been
numerically analyzed in \cite{tree1,tree2} at tree level.
Electroweak corrections to the widths of $H^{\pm}\to
\tilde\chi^{\pm}_{i} \ch^0_m$ due to one-loop exchanges of the third
generation quarks and squarks were recently calculated in~\cite{Yi}.
The one-loop corrections, involving fermions and sfermions, to the
invisible width of  $(h^0, H^0, A^0)\to\ch^0_1  + \ch^0_1$ have been
calculated in the higgsino limit of  $\ch^0_1,\,(|\mu |\ll M_1
,M_2)$ in~\cite{higgsinolike}, and in the gaugino limit of
$\ch^0_1,\,(|\mu |\gg M_1 ,M_2)$ very recently
in~\cite{Djouadi-Drees}. (Here $M_1$ and $M_2$ are the $U(1)$ and
$SU(2)$ gaugino mass parameters, respectively, and $\mu$ is the
higgsino mass parameter.) In these limiting cases, the wave-function
corrections can be neglected and no renormalization is necessary.
The couplings of $(h^0,H^0, A^0)$ to $\ch^0_1$ $\ch^0_1$ also enter
in the neutralino-quark interaction~\cite{Djouadi-Drees}, a process
which is very important for the dark matter
search~\cite{dark1,dark2}, where one looks for the elastic
scattering of neutralinos $\ch^0_1$ off nuclei in a detector.
Moreover, since the decays (\ref{eq:h0ch1ch1},\ref{eq:Hkchmchn}) are
generated by gaugino-higgsino-Higgs boson couplings at tree level,
they can be also useful to probe the components of the neutralinos,
complementary to the pair production process
$e^+e^-\hspace{-3pt}\rightarrow\ch^0_m\ch^0_n$~\cite{neuproduction}.

In this paper, we present the one-loop corrections to the widths
of the decays~(\ref{eq:h0ch1ch1}) and~(\ref{eq:Hkchmchn}) due to
the exchange of all fermions (quarks and leptons) and their
superpartners (sfermions).

The decays (\ref{eq:h0ch1ch1}) and (\ref{eq:Hkchmchn}) are
particularly interesting because the calculation of their
radiative corrections requires corrections to the neutralino mass
matrix and mixing matrix in addition to the conventional
wave-function and vertex corrections with counter terms. The
one-loop corrections to the neutralino mass and mixing matrix in
the on-shell renormalization scheme were already worked out in
\cite{chmasscorr} and they will be used here.\\
Related to these decays are the decays of neutralinos into Higgs bosons,
\begin{equation}
 \ch^0_m \to (h^0, H^0, A^0)  + \ch^0_n \, .
 \label{eq:chidecay}
\end{equation}
These decays are also important as they occur in the cascade
decays of gluinos and/or squarks, $\tilde{g}\to
q\bar{q}\tilde\chi^0_m$ and $\tilde{q}\to q\tilde\chi^0_m$, with
$\tilde\chi^0_m$ then decaying according to (\ref{eq:chidecay}).
The decays (\ref{eq:chidecay}) with a real Higgs boson emission
\cite{neutree1,neutree2} as well as three-body decays due to an
off-shell Higgs boson \cite{neutree3} have been studied at tree
level. In this paper, we also present the formulae for the decays
(\ref{eq:chidecay}) including
the one-loop corrections.

\section{Tree-level widths}

Throughout this paper, we will use the notations $m_{\ch^0_i}
\equiv m_i$ and $H_k^0 \equiv \{h^0, H^0, A^0, G^0 \}$. In a
non-unitary gauge we have the ghost $G^0$.
The momenta are assigned as $(k = 1,2,3$; $m,n=1,\ldots,4)$
\begin{equation}
  H_k^0(p) \to \ch^0_m (k_1) + \ch^0_n (k_2) \, .
\label{eq:Hk0neu}
\end{equation}
All couplings are given in the Appendix~\ref{sec:appA} (or it is
referred to previous works).

The tree-level widths for a neutral Higgs decaying into two neutralinos
is \cite{tree1}
\begin{eqnarray}
 && \hspace{-1cm}
 \Gamma^{\rm tree}_H = \Gamma^{\mbox{\footnotesize
tree}}(H_{k}^0 \to \ch^{0}_{m}\, \ch^{0}_{n}) = \nonumber\\[2mm]
 && \hspace{-0.5cm}
 \frac{g^{2}}{8 \pi\, m^{3}_{H_{k}^0}\,
(1 + \delta_{mn})} \, \kappa(m^{2}_{H_{k}^0},m^{2}_{m},m^{2}_{n}) \,
|F^0_{mnk}|^{2}\, \left[ m^{2}_{H_{k}^0} - m^{2}_m - m^{2}_n
 - 2 (-1)^{\delta_{k3}}\, m_m m_n  \right]\, ,
\label{eq:gammatree}
\end{eqnarray}
with $\kappa(x,y,z)\equiv ((x-y-z)^2-4yz)^{1/2}$. The couplings
$F^0_{mnk}$ are given in the Appendix,
eqs.~(\ref{eq:F0lag}-\ref{eq:dudd}).

For the decay of a neutralino into a lighter one and $H_k^0$, we
get \cite{neutree1}
\begin{eqnarray}
 && \hspace{-1.5cm} \Gamma^{\rm tree}_{\ch^0} = \Gamma^{\mbox{\footnotesize
tree}}(\ch^{0}_m \to
 H_{k}^0\, \ch^{0}_n) = \nonumber\\[2mm]
 && \hspace{0.5cm}
 \frac{g^{2}}{16 \pi\, m^{3}_m} \, \kappa(m^{2}_m, m^{2}_{H_{k}^0}, m^{2}_n)
\, |F^0_{mnk}|^{2}\,\left[ m^{2}_m - m^{2}_{H_{k}^0} + m^{2}_n + 2
(-1)^{\delta_{k3}}\, m_m m_n \right]\, .
%\label{}
\end{eqnarray}
In our convention, the $4 \times 4$ neutralino mixing matrix $Z$,
which diagonalizes the neutralino mass matrix $Y$, is real. Therefore,
the neutralino mass parameters $m_m$ and $m_n$ can be positive or negative.

\section{One-loop corrections}
We calculate the one-loop corrections to the amplitudes of the
decays (\ref{eq:Hk0neu}) stemming from fermion and sfermion
exchange. The renormalization is done in the on-shell scheme. All
one-, two-, and three-point functions \cite{thooft} used for
calculating the loop integrals are given in the convention
\cite{denner}.\\

The correction to the coupling $F^0_{mnk}$ is
\begin{equation}
  \label{eq:Fren}
F^{0\, {\rm corr.}}_{mnk} = F^0_{mnk} + \Delta F^0_{mnk}\, ,
\end{equation}
with the ultraviolet (UV) finite one-loop correction
\begin{equation}
\label{eq:DFmnk}
 \Delta F^0_{mnk} = \sum_{\rm flavors}\, N_c^f\,
\left(\d F^{0\,(v)}_{mnk} + \d F^{0\,(w)}_{mnk} +
 \d F^{0\,(c)}_{mnk} \right)\, ,
\end{equation}
with the color factor $N_c^f = 1$ for (s)lepton and $N_c^f = 3$
for (s)quark exchange. $\sum_{\rm flavors}$ stands for the
summation over all (s)fermion flavors, e.~g.~(top, stops),
(bottom, sbottoms), (tau, staus), etc.. For convenience, the color
factor $N_c^f$ is given only in the total correction term
eq.~(\ref{eq:DFmnk}).

In our convention, both $F^0_{mnk}$ and $\Delta F^0_{mnk}$ are real.
Therefore, the corrected widths can be written as
\begin{equation}
  \label{eq:Gcorr}
\Gamma^{\rm corr.}_p =  \Gamma^{\rm tree}_p \, \left(1 + \frac{\Delta
F^0_{mnk}}{F^0_{mnk}} \right)^2\, ,
\end{equation}
with the decaying particle $p = H_k^0$ or $\ch^0_m$.\\

\begin{figure}
 \begin{center}
 \mbox{\resizebox{14cm}{!}{\includegraphics{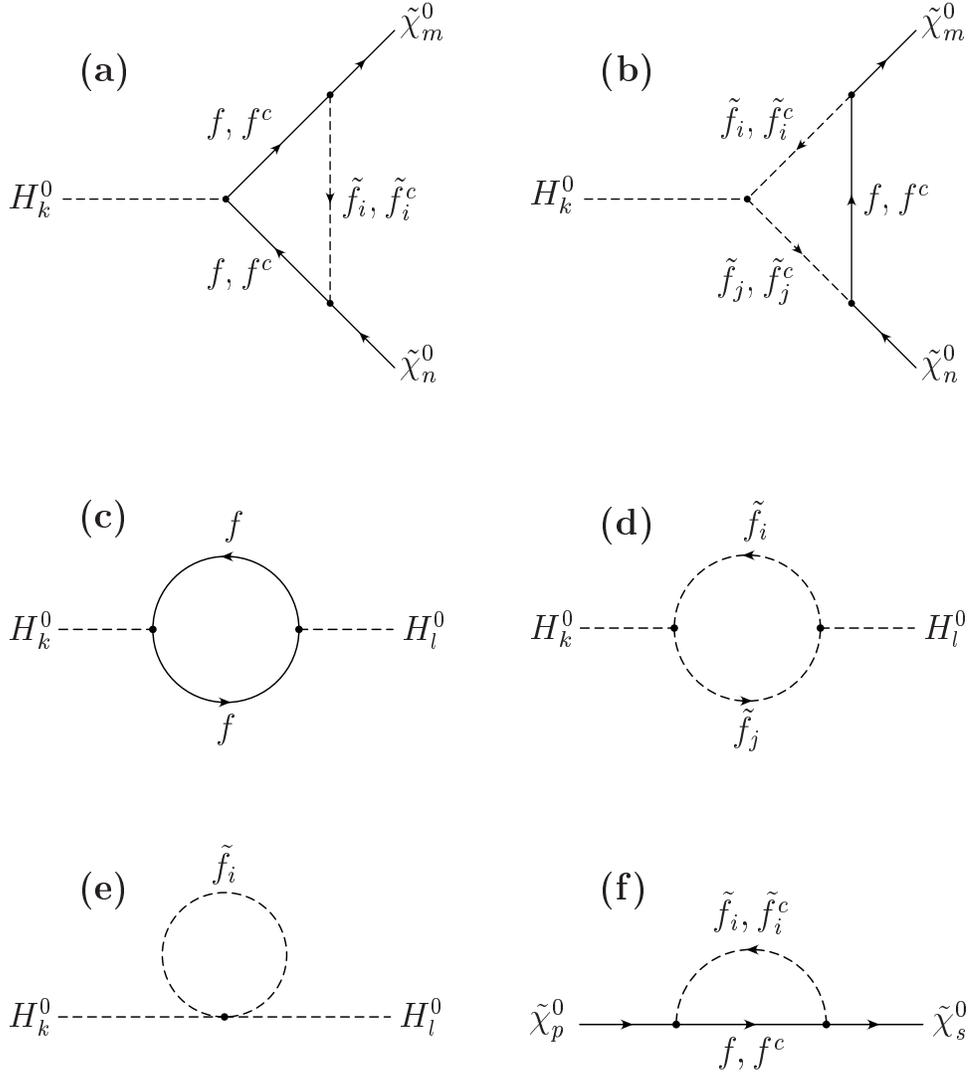}}}
  \end{center}
  \caption[feynman]{One-loop Feynman graphs with fermion and
  sfermion exchange contributing
  to the neutral Higgs boson-neutralino-neutralino decay amplitude.
  The superscript ``$c$'' denotes the charge conjugated states.
  \label{fig:feynman}}
 \end{figure}

The {\bf vertex correction} stems from the two diagrams shown in
Figs.~\ref{fig:feynman}a and \ref{fig:feynman}b. Because of the
Majorana nature of the neutralinos the charge conjugated
(s)fermion fields denoted by the superscript ``$c$'' can also
circulate in the loop.\\ For $h^0$ and $H^0$ ($a = 1,2$) we have
\begin{eqnarray}
 &&\hspace*{-1.1cm} g\, \d F^{(v)}_{mna} =
 \frac{s^f_a}{(4 \pi)^2}
  \sum_{i=1}^2
 \Bigg\{m_f\Big(a^{\tilde{f}}_{im}a^{\tilde{f}}_{in}+b^{\tilde{f}}_{im}
 b^{\tilde{f}}_{in}\Big) \bigg[\Big(m_{m} + m_{n}\Big) C_0^{i} +
 2\Big(m_{m} C_1^{i} + m_{n}C_2^{i}\Big)\bigg] +
 \nonumber\\
 &&\hspace*{-1.1cm}+
 \Big(a^{\tilde{f}}_{im}b^{\tilde{f}}_{in}+
  a^{\tilde{f}}_{in}b^{\tilde{f}}_{im}\Big)
    \bigg[\Big(m_f^2+m_{\sf_i}^2+m_{m}m_{n}\Big)C_0^{i}+
   \Big(m_{m}+m_{n}\Big)\Big(m_{m}C_1^{i}+m_{n}C_2^{i}\Big)
   +B_0^{}\bigg]
  \Bigg\}
  \nonumber\\
 &&\hspace*{-1.1cm}+
 \frac{1}{(4 \pi)^2}
  \sum_{i,j\,= 1}^2 G_{ija}^{\tilde{f}}\,
 \Bigg[\Big(a^{\tilde{f}}_{im}b^{\tilde{f}}_{jn}+a^{\tilde{f}}_{jn}
 b^{\tilde{f}}_{im}\Big)m_f C_0^{ij}-
 \Big(a^{\tilde{f}}_{im}a^{\tilde{f}}_{jn}+
 b^{\tilde{f}}_{im}b^{\tilde{f}}_{jn}\Big)
  \Big(m_{m} C_1^{ij} + m_{n} C_2^{ij}\Big)
 \Bigg]
   \, .  \label{eq:dvgFSa}
\end{eqnarray}
For $A^0$ the vertex correction reads
\begin{eqnarray}
&&\hspace*{-1.1cm} g\,
\d F^{(v)}_{mn3} =
  \frac{i\,s^f_3}{\left(4\pi\right)^2}
   \sum_{i=1}^2
  \Bigg\{m_f
 \Big(a^{\tilde{f}}_{im}a^{\tilde{f}}_{in}+
 b^{\tilde{f}}_{im}b^{\tilde{f}}_{in}\Big)
 \Big(m_{m}+m_{n}\Big)\,C_0^{i}+
 \nonumber\\
 &&\hspace*{-1.1cm} +
 \Big(a^{\tilde{f}}_{im}b^{\tilde{f}}_{in}+
 a^{\tilde{f}}_{in}b^{\tilde{f}}_{im}\Big)\bigg[
 \Big(m_{m} m_{n}-m_{\sf_i}^2+m_f^2\Big)C_0^{i}+
 \Big(m_{n}-m_{m}\Big)\Big(m_{m}C_1^{i}-m_{n}C_2^{i}\Big)
 -B_0\bigg]
 \Bigg\}\nonumber\\
 &&\hspace*{-1.1cm}+
 \frac{i}{(4 \pi)^2}\,\sum_{i,j\,= 1}^2 G_{ij3}^{\tilde{f}}\,
 \Bigg[\Big(a^{\tilde{f}}_{im}b^{\tilde{f}}_{jn}-a^{\tilde{f}}_{jn}
 b^{\tilde{f}}_{im}\Big) \,m_f \,C^{ij}_0 +
 \Big(a^{\tilde{f}}_{im}a^{\tilde{f}}_{jn}-
 b^{\tilde{f}}_{im}b^{\tilde{f}}_{jn}\Big)\,
  \Big(m_{m} C^{ij}_1 - m_{n} C^{ij}_2\Big)
 \Bigg] . \,\label{eq:dvgFSc}
 \end{eqnarray}
\noindent The abbreviations $B_0 = B_0(m_{H^0_k}^2,m_f^2,m_f^2)$, $C^i_{..} =
 C_{..}(m_m^2,m_{H^0_k}^2,m_n^2,m_{\sf_i}^2,m_f^2,m_f^2)$, and
 $C^{ij}_{..}= C_{..}(m_m^2,m_{H^0_k}^2,
 m_n^2,m_f^2,m_{\sf_i}^2,m_{\sf_j}^2)$ have been used.

The {\bf wave-function correction} is given by
 \begin{equation}
 \delta{F}_{mnk}^{0\,(w)}\,=\, \frac{\,1}{\,2}\,\bigg[\,
\delta{Z}^{H^0}_{lk}{F}_{mnl}^{0}+ \delta{Z}_{qm}^{\ch}{F}_{qnk}^{0}
+\delta{Z}_{qn}^{\ch}{F}_{mqk}^{0}\bigg]
 \,,
 \label{eq:dFw}
 \end{equation}
with the implicit summations over $l = 1,2$ for k = 1 or 2, $l =
3,4$ for $k = 3$, and $q = 1,\ldots,4$. $\delta{Z}_{}^{\ch}$ are
the wave-function constant terms for the neutralinos given
in~(\ref{eq:dZchpp}),~(\ref{eq:dZchps}). The wave-function
constant terms for the Higgs bosons $\delta{Z}^{H^0}$ are
\begin{eqnarray}
 \label{eq:dZHkk}
 \delta{Z}^{H^0}_{kk}&=&
  -\;{\rm Re}\,\dot{\Pi}^{H^0}_{kk}(m_{H_k^0}^2)\, , \\ \label{eq:dZHlk}
\delta{Z}^{H^0}_{kl}&=& \frac1{m^2_{H_k^0}-m^2_{H_l^0}}\, {\rm
Re}\,\bigg\{\Pi^{H^0}_{kl}(m^2_{H_l^0}) - \Pi^{H^0}_{kl}(m^2_{H_k^0})\bigg\} \,
, \hspace{20mm}l\neq{k}\,,
\end{eqnarray}
with $k,l = 1,2$ for the system $(h^0, H^0)$ and $k,l = 3,4$ for
$(A^0, G^0)$. Eq.~(\ref{eq:dZHlk}) has been symmetrized with
respect to $(k,l)$. This is due to the on-shell renormalization of
the Higgs mixing angle $\alpha$ ($k,l=1,2$) or $\beta$
($k,l=3,4$). In this scheme (\cite{chmasscorr}, extending
\cite{earlier} for quark and lepton mixing) the counter terms for
the mixing angles are determined by the requirement that they
cancel the antisymmetric parts of the wave-function corrections.

The decays of $A^0$ are a little complicated by the contribution
of the $A^0-Z^0$ mixing in addition to the $A^0-G^0$ mixing in
eq.~(\ref{eq:dZHlk}). Moreover, both depend on the gauge parameter~$\xi$.
However, the sum of these two contributions is independent
of $\xi$, as it is shown in Appendix~C. Here we work in the $\xi = 0$
(Landau) gauge, where the contribution of the $A^0-Z^0$ mixing
vanishes, and use (\ref{eq:dZHlk}) with $m_{H^0_4}=0$. The
resulting on-shell $\tan\beta$ agrees with the one defined by the
$A^0$--$Z^0$ mixing \cite{pokorski,dabelstein, chmasscorr}.

The Higgs self-energy contributions due to fermions and sfermions are
written as
\begin{equation}
\Pi^{H^0}_{kl}(k^2) = \Pi^{H^0\,(a)}_{kl}(k^2) +
\Pi^{H^0\,(b)}_{kl}(k^2) + \Pi^{H^0\,(c)}_{kl} + T_{kl}\,.
\label{eq:TotalHiggsSE}
\end{equation}
The fermion contribution $\Pi^{H^0\,(a)}$ (Fig.~\ref{fig:feynman}c) is
 \begin{eqnarray}
 \Pi^{H^0\,(a)}_{ab}(k^2)&=&\frac{2\,s^f_a\,s^f_b}{\left(4\pi\right)^2}\,
 \bigg[\Big(k^2-4m_f^2\Big)\,B_0(k^2,m_f^2,m_f^2)-2A_0(m_f^2)\bigg]\,,
  \label{eq:PiHa1}
\\
 \Pi^{H^0\,(a)}_{cd}(k^2)&=&-\frac{2\,s^f_c\,s^f_d}{\left(4\pi\right)^2} \,
 \bigg[k^2\,B_0(k^2,m_f^2,m_f^2)-2A_0(m_f^2)\bigg]\,,
\label{eq:PiHa2}
  \end{eqnarray}
with $a,b = 1,2$ and $c,d = 3,4$. The sfermion contributions
$\Pi^{H^0\,(b)}$ (Fig.~\ref{fig:feynman}d) and
$\Pi^{H^0\,(c)}$ (Fig.~\ref{fig:feynman}e) are
\begin{eqnarray}
 \Pi^{H^0\,(b)}_{kl}(k^2) & = &\frac{1}{\left(4\pi\right)^2}\;
 \sum_{i,j=1}^2
 G^{\sf}_{jik}\, G^{\sf}_{ijl}\;
 B_0(k^2,m_{\sf_i}^2,m_{\sf_j}^2)\,,
 \label{eq:PiHb}
 \\
 \Pi^{H^0\,(c)}_{kl}& = & \forpi \;
 \sum_{i=1}^2
  \bigg[h_f^2\,c_{kl}^\sf\,+\,g^2d_{kl}\,e_{ii}^\sf\bigg]
  A_0(m_{\sf_i}^2)\,,
  \label{eq:PiHc}
\end{eqnarray}
where $k,l = 1,2$ or $3,4$. $T_{kl}$ in
eq.~(\ref{eq:TotalHiggsSE}) represent momentum-independent
contributions from the tadpole shifts~\cite{pokorski,dabelstein}
and leading higher-order corrections. We include the latter by the
renormalization group improvement as in
Ref.~\cite{twopointfunctions}. Since the zero-momentum
contribution $\Pi^{H^0}_{kl}(0)$, including $T_{kl}$, is very
large it is often resummed as in Refs.~\cite{dabelstein,carena}. In
practice, we calculate the effective $m_{H^0_a}(a=1,2)$ and
$\alpha$ obtained from the effective mass matrix, which includes
the $\Pi^{H^0}_{ab}(k^2=0)$ contribution with $m_{A^0}$, $\tb$,
and the (s)quark parameters, and regard them as the lowest-order
parameters. If one is replacing $\alpha$ in all the previous formulae
with the effective one, the self energies $\Pi^{H^0}_{kl}(k^2)$ in the
wave-function correction and $\delta\alpha$ must be replaced by
$\Delta\Pi^{H^0}_{kl}(k^2)=\Pi^{H^0}_{kl}(k^2)-\Pi^{H^0}_{kl}(0)$.
Nevertheless, the form of their sums eqs.~(\ref{eq:dZHkk},\ref{eq:dZHlk}) is
not affected by the elimination of
$\Pi^{H^0}_{kl}(0)$.

The neutralino wave-function terms read
\begin{eqnarray}
\hspace*{-1cm}  \delta{Z}^{\ch}_{pp} & = & - \;{\rm Re}\,
 \bigg\{\Pi^{\ch}_{pp}(m_{p}^2)+ 2m_{p}\,
 \Big[m_{p}\dot{\Pi}^{\ch}_{pp}(m_{p}^2)+
 m_f\dot{\Pi}^{\ch_S}_{pp}(m_{p}^2)\Big]\bigg\}\, ,
 \label{eq:dZchpp}
 \\
 %[3mm]
 \label{eq:dZchps}
\hspace*{-1cm}   \delta{Z}^{\ch}_{ps}&=&\frac1{m_{p}-m_{s}}\;
 {\rm Re}\bigg\{
 m_{s}\left[\Pi^{\ch}_{ps}(m_{s}^2) - \Pi^{\ch}_{ps}(m_{p}^2)\right] + m_f
\left[\Pi^{\ch\,S}_{ps}(m_{s}^2) - \Pi^{\ch\,S}_{ps}(m_{p}^2) \right]\bigg\} \,,
\end{eqnarray}
$p\neq s$. As before, $\delta{Z}^{\ch}_{ps}$ in (\ref{eq:dZchps})
has been symmetrized by subtracting the counter term for the
rotation matrix $Z$ of the neutralinos \cite{chmasscorr}.
The neutralino self-energies due to
the sfermion-fermion loop (Fig.~\ref{fig:feynman}f ) are
\begin{eqnarray}
\label{eq:neuself} \Pi^\ch_{ps}(k^2) & = & - \frac{1}{(4\pi)^2}\,
\sum_{i=1}^2(a^{\tilde{f}}_{ip} a^{\tilde{f}}_{is}+ b^{\tilde{f}}_{ip}
b^{\tilde{f}}_{is})\, B_1(k^2,m_f^2,m_{\sf_i}^2)\, ,
\\
\label{eq:neuselfS} \Pi^{\ch\,S}_{ps}(k^2) & = & \frac{1}{(4\pi)^2}\,
\sum_{i=1}^2(a^{\tilde{f}}_{ip} b^{\tilde{f}}_{is}+ a^{\tilde{f}}_{is}
b^{\tilde{f}}_{ip})\, B_0(k^2,m_f^2,m_{\sf_i}^2)\, ,
\end{eqnarray}
 see also \cite{twopointfunctions}.

We need the {\bf counter term} for the couplings $gF^0_{mnk}$,
which is a function of the gauge couplings $g$, $g'$, the Higgs
boson mixing angle $\alpha$ (for $k=1,2$) or $\beta$ (for $k=3$),
and the neutralino rotation matrix $Z$, as shown in
eq.~(\ref{eq:F0tree}). The counter terms for ($\alpha$, $\beta$)
and $Z$ in the on-shell scheme \cite{chmasscorr} are already
included in the wave-function corrections (\ref{eq:dZHlk}) and
(\ref{eq:dZchps}), respectively. The remaining counter term of $\d
g$ and $\d g'$ is, after being
absorbed into the correction to $F^0_{mnk}$,
\begin{eqnarray}
 \d F^{0\, (c)}_{mnk} &=& \sum_{x = 3,4}\,
 d^x_k \left[\left(Z_{mx}Z_{n2}+Z_{m2}Z_{nx}\right)\, \frac{\d g}{g}
 -  \left(Z_{mx}Z_{n1}+Z_{m1}Z_{nx}\right)\, \tan\theta_W\, \frac{\d
 g'}{g'}\right]
 \,.
 \label{eq:dgF0c}
\end{eqnarray}
We fix the electroweak gauge boson sector by $m_Z$, $m_W$, and $e$.
One gets from the relations $g=e/s_W$, $g'=e/c_W$, and $c_W={m_W}/{m_Z}$
($c_W\equiv \cos\theta_W$, $s_W \equiv \sin\theta_W$)~\cite{sirlin,denner}
 \begin{equation}
  \frac{\delta g}{g}=  \frac{\delta e}{e} +\frac{c_W^2}{2\,s_W^2}
 \left(\frac{\delta m_W^2}{m_W^2}-\frac{\delta m_Z^2}{m_Z^2}\right)\, ,
 \hspace{1.5cm}
 \frac{\delta g'}{g'}= \frac{\delta e}{e} -
 \frac12\left(\frac{\delta m_W^2}{m_W^2}-
 \frac{\delta m_Z^2}{m_Z^2}\right)\,.
 \label{eq:dggdgpgp}
 \end{equation}
The formulae for $\delta m_W$ and $\delta m_Z$ can be also found in
\cite{chmasscorr} and for $\d e/e$ in the Appendix~\ref{sec:appB}.

Now all parts are given which are needed in order to calculate the
(UV finite) one-loop contribution to the neutral
Higgs boson-neutralino-neutralino coupling, eq.~(\ref{eq:DFmnk}). The
vertex correction part $\d F^{0\,(v)}_{mnk}$ is given by
eqs.~(\ref{eq:dvgFSa}) and (\ref{eq:dvgFSc}), the wave-function
correction term $\d F^{0\,(w)}_{mnk}$ by eq.~(\ref{eq:dFw}), and $\d
F^{0\,(c)}_{mnk}$ by eq.~(\ref{eq:dgF0c}).

Further, one has to note that the on-shell masses and the mixing of
the neutralinos are not independent of each other. In fact, when
the gauge and Higgs boson sectors are fixed, the neutralino sector
is determined by three free parameters only. Here we follow the
method given by \cite{chmasscorr}: The on-shell mass parameters
$M$ and $\mu$ are defined as the elements of the on-shell mass
matrix $X$ of charginos, and the on-shell mass parameter $M'$ is
defined as the element of the on-shell mass matrix $Y$ of
neutralinos . The finite correction $\Delta Y=Y-Y^{\rm tree}$,
where $Y^{\rm tree}$ is the tree-level mass matrix in terms of the
on-shell parameters $(M,\mu,M',m_Z,\sin\theta_W,\tan\beta)$, is
calculated by eqs.~(42--51) in \cite{chmasscorr}. The one-loop
corrected on-shell masses $m_i$ and mixing matrix
$Z=Z^{\rm tree}+\Delta Z$ are then obtained by diagonalizing $Y$.
For a proper treatment of the loop corrections, the resulting
shifts of the masses and the mixing matrix from the tree-level values
have to be taken into account.

\section{Numerical results}

For simplicity, we will take in the following (if not specified
otherwise) for the soft breaking sfermion mass parameters of the
first, second and third generation $M_{\tilde Q_{1,2}}=M_{\tilde
U_{1,2}}=M_{\tilde D_{1,2}}=M_{\tilde L_{1,2}}=M_{\tilde
E_{1,2}}=M_{\tilde Q_{3}}=\frac{10}9M_{\tilde U_{3}}
=\frac{10}{11}M_{\tilde D_{3}}=M_{\tilde L_{3}}=M_{\tilde E_{3}} =
M_{\tilde Q}=500$~GeV  and for the trilinear couplings
$A_t=A_b=A_{\tau}=A=300$~GeV. We take $m_Z=91.2$~GeV,
$m_W=80$~GeV, $m_{A^0}=500$~GeV, $m_t=175$~GeV, $m_b=5$~GeV, and
$m_{\tau}=1.8$~GeV. Masses of all other SM fermions are neglected.
We use the GUT relations $M'=\frac53\tan^2{\theta_W}\,M$ and for
the gluino mass
$m_{\tilde{g}}=({\alpha_S}/{\alpha_{EW}})\sin^2{\theta_W}\,M$. The
other input parameters are $\{\,\tan{\beta},\,M,\,\mu\}$ (all as
on-shell parameters). For the values of the Yukawa couplings of
the $3^{\rm rd}$ quark generation ($h_t$, $h_b$), we take the
running ones at the scale of the decaying particle mass.

In our numerical analysis we have discussed four cases: the
tree-level width, the corrections (\ref{eq:Fren}--\ref{eq:Gcorr})
with the tree-level $Z$ and $m_i$ (``conventional correction''),
the corrections (\ref{eq:Fren}--\ref{eq:Gcorr}) with the one-loop
corrected $Z$ and tree-level $m_i$ (``conventional $+$ $\Delta Z$
correction''), and the corrections (\ref{eq:Fren}--\ref{eq:Gcorr})
with the one-loop corrected $Z$ and one-loop corrected $m_i$ (full
correction). The ``conventional correction'' corresponds to the
correction to the gaugino-higgsino-Higgs boson coupling,
``conventional $+$ $\Delta Z$ correction'' includes the correction
to the neutralino components, and the correction due to the shift
of $m_i$ is added in the full correction.

\begin{figure}[h!]
 \begin{center}
 % \vspace*{-10mm}
 \hspace{-8mm}
 \mbox{\resizebox{84mm}{!}{\includegraphics{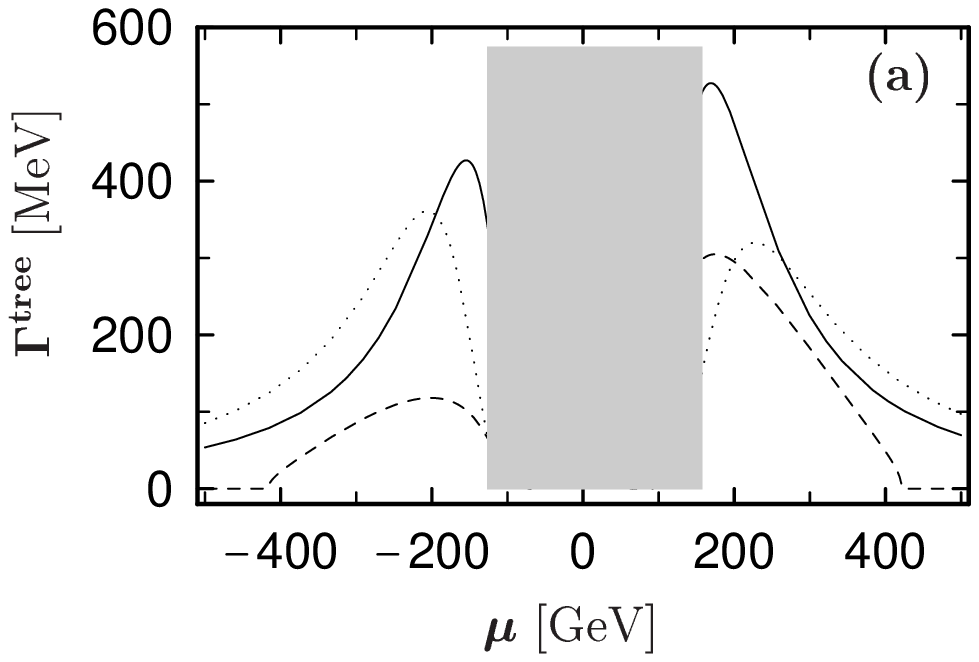}}}
 \hspace{-3mm}
 \mbox{\resizebox{82mm}{!}{\includegraphics{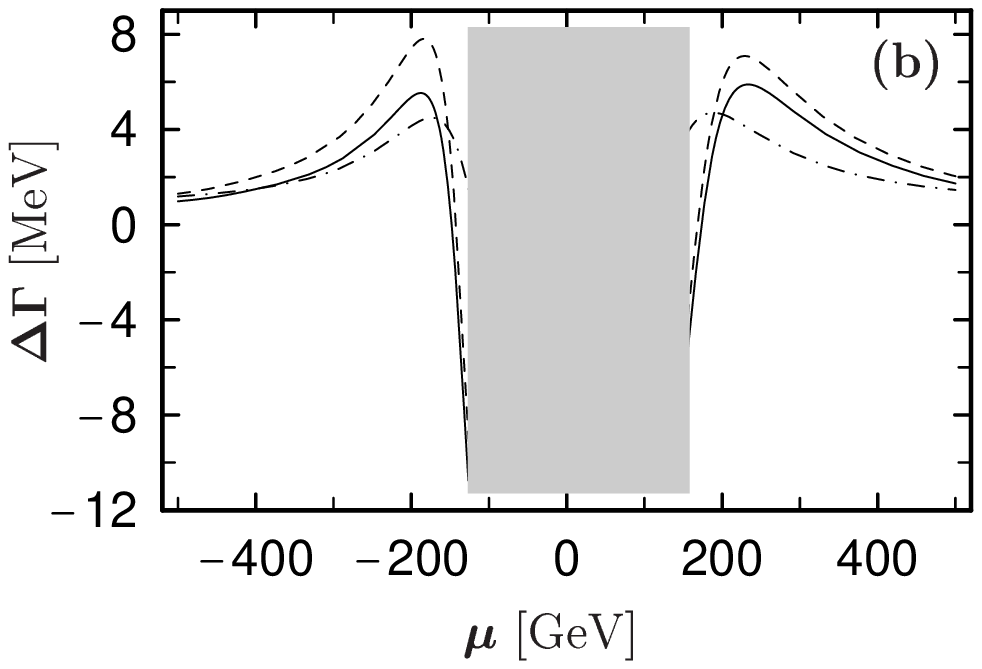}}}
 \\
 \hspace{-8mm}
 \mbox{\resizebox{81mm}{!}{\includegraphics{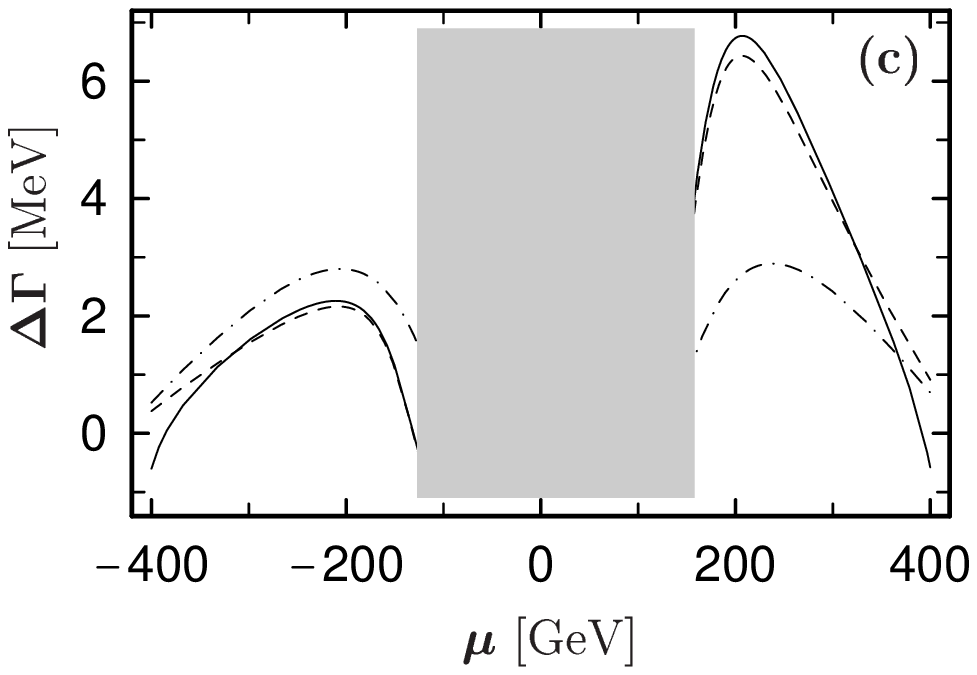}}}
 \hspace{-2mm}
 \mbox{\resizebox{81mm}{!}{\includegraphics{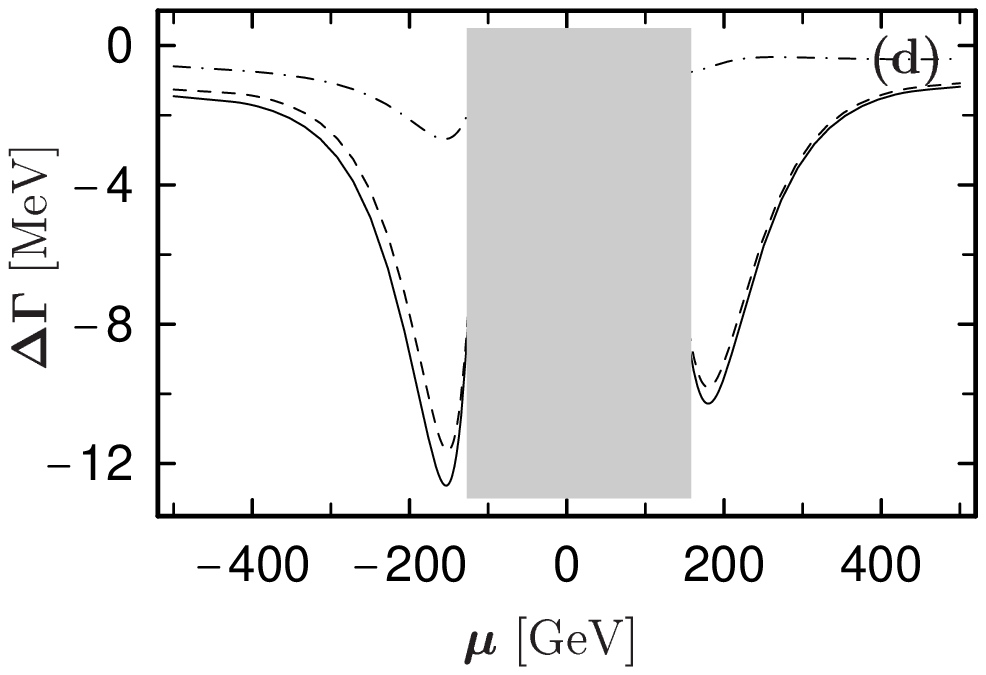}}}
 \hspace{-10mm}
 \vspace{-0mm}
 \caption[fig5]
 {Tree-level widths (a) of the decays
$H^0\to\ch^0_1+\ch^0_2$~(solid), $H^0\to\ch^0_1+\ch^0_3$~(dashed) and
$H^0\to\ch^0_2+\ch^0_2$~(dotted) and corrections to the
widths of these decays (b), (c), and (d), respectively, as a function of $\mu$ for
$\tan{\beta}=10$ and $M=150$~GeV. The full, dashed, dash-dotted
line corresponds to the full, ``conventional + $\Delta{Z}$'', and ``conventional'' correction.
 The grey areas are excluded by the bounds $m_{\ch^\pm_1}\geq100$~GeV,
 $m_{h^0}>92$~GeV.}
 \label{fig:mudependence}
 \end{center}
 \vspace{-5mm}
 \end{figure}

In Fig.~\ref{fig:mudependence}a we show, as a function of $\mu$,
the tree-level widths of $H^0\to\ch^0_1+\ch^0_2$,
$H^0\to\ch^0_1+\ch^0_3$ and $H^0\to\ch^0_2+\ch^0_2$, respectively,
for $\tan{\beta}=10$ and $M=150$~GeV. The $H^0$ mass is
$m_{H^0}\sim500$ GeV.  The widths vary with the gaugino and
higgsino components of the various neutralino states.
Fig.~\ref{fig:mudependence}b exhibits the corrections to the width
of $H^0\to\ch^0_1+\ch^0_2$: The ``conventional'', ``conventional
$+\Delta Z$'', and full corrections are shown. One can see that,
compared to the ``conventional'' correction, the corrections by
the shifts $\Delta Z$ and $m_i$ cannot be neglected.
Figs.~\ref{fig:mudependence}c and \ref{fig:mudependence}d show the
corrections to the widths of $H^0\to\ch^0_1+\ch^0_3$ and
$H^0\to\ch^0_2+\ch^0_2$, respectively. While the ``conventional''
correction is dominant for $\mu<0$ in
Fig.~\ref{fig:mudependence}c, the ``$\Delta Z$'' correction is
dominant in Fig.~\ref{fig:mudependence}d.
The full corrections amount to several \%.

\begin{figure}[h!]
 \begin{center}
 % \vspace*{-10mm}
 \hspace{-8mm}
 \mbox{\resizebox{84mm}{!}{\includegraphics{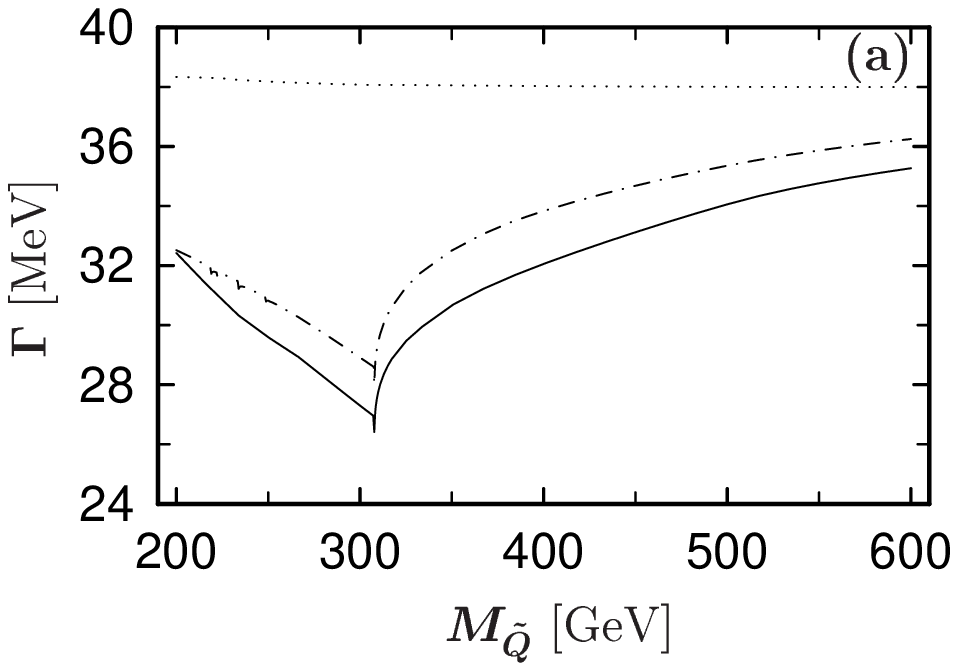}}}
 \hspace{-3mm}
 \mbox{\resizebox{84mm}{!}{\includegraphics{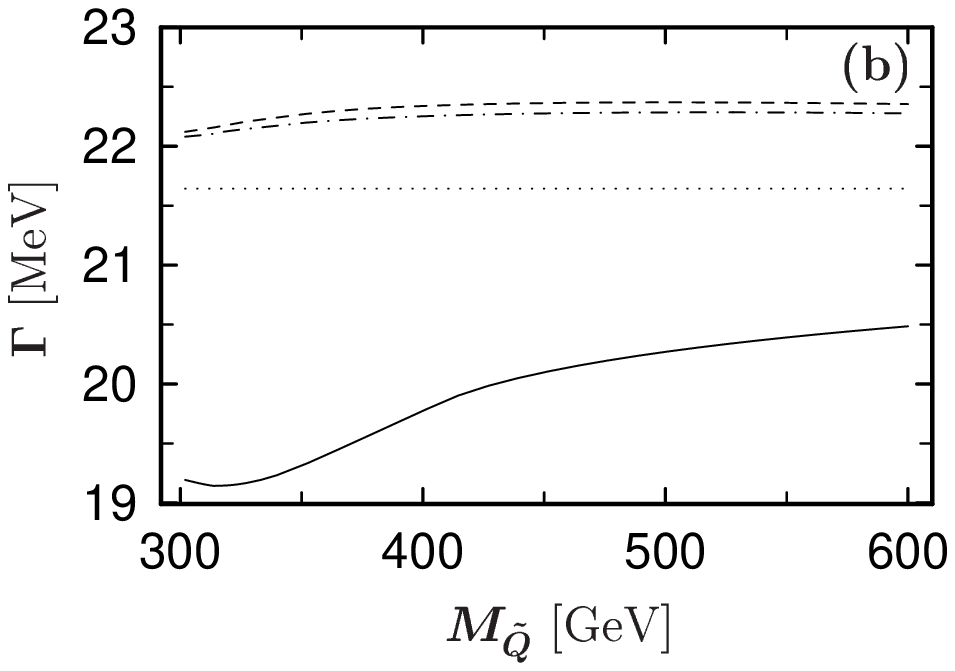}}}
 \hspace{-10mm}
 \vspace{-0mm}
 \caption[fig5]
 {The widths of the decays $H^0\to\ch^0_1+\ch^0_2$~(a) and
 $A^0\to\ch^0_1+\ch^0_3$~(b) as a function of $M_{\tilde{Q}}$. The
 dotted line corresponds to the tree-level width, the
 dash-dotted, dashed, and solid line corresponds to the
 ``conventional'', ``conventional + $\Delta{Z}$'', and full correction,
 respectively. The parameters are $\tan{\beta}=10$, $M=500$~GeV,
 and $\mu=150$~GeV~(a) and $\tan{\beta}=50$ and $M=\mu=300$~GeV~(b). }
 \label{fig:MsQdependence}
 \end{center}
 \vspace{-5mm}
 \end{figure}

In Fig.~\ref{fig:MsQdependence}, we show the tree and corrected
widths of (a) $H^0\to\ch^0_1+\ch^0_2$ decay with $\tan\beta=10$,
$M=500$~GeV and $\mu=150$~GeV, and those of (b)
$A^0\to\ch^0_1+\ch^0_3$ decay with $\tan\beta=50$ and
$M=\mu=300$~GeV, as functions of $M_{\tilde{Q}}$. In
Fig.~\ref{fig:MsQdependence}a, the decay is suppressed due to the
small gaugino components of $\ch^0_1$ and $\ch^0_2$. The
``conventional$+\Delta Z$'' correction is close to the full
correction and therefore not shown here. We see that the
``conventional'' correction is dominant. In contrast, the
``$\Delta m$'' correction in Fig.~\ref{fig:MsQdependence}b is
large and negative (up to $-16$\%), which dominates over the
positive ``conventional'' correction (up to $+4$\%). This is
because the decay in Fig.~\ref{fig:MsQdependence}b is
kinematically suppressed and sensitive to the shift of $m_i$. We
note that the sfermion loop corrections do not decouple in large
$M_{\tilde{Q}}$ limit, due to the supersymmetry breaking
corrections~\cite{superoblique} to the gaugino-higgsino-Higgs
boson couplings.

\begin{figure}[h!]
 \begin{center}
 % \vspace*{-10mm}
 \hspace{-8mm}
 \mbox{\resizebox{84mm}{!}{\includegraphics{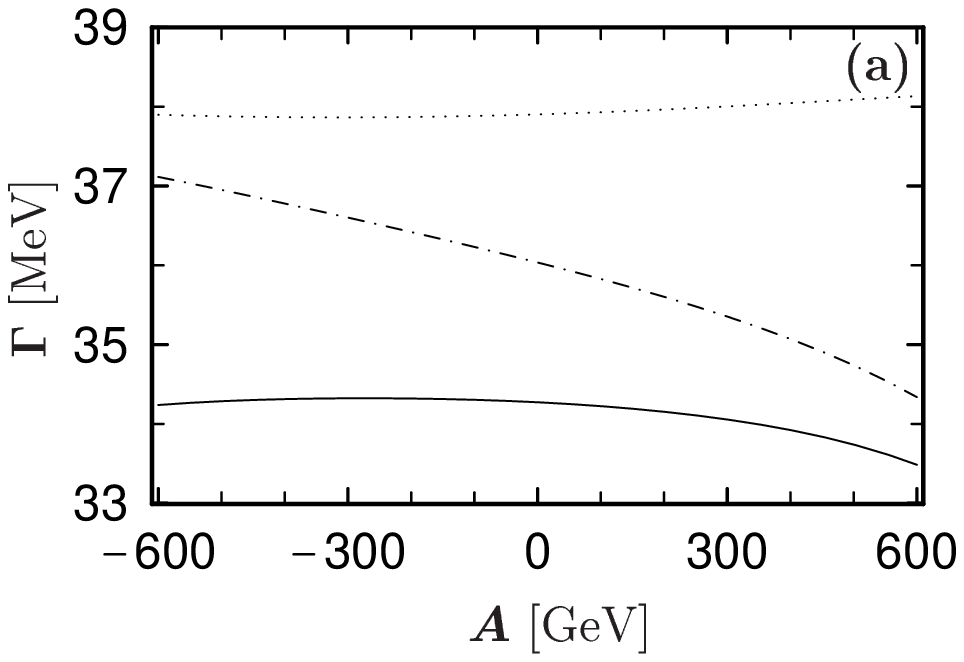}}}
 \hspace{-3mm}
 \mbox{\resizebox{84mm}{!}{\includegraphics{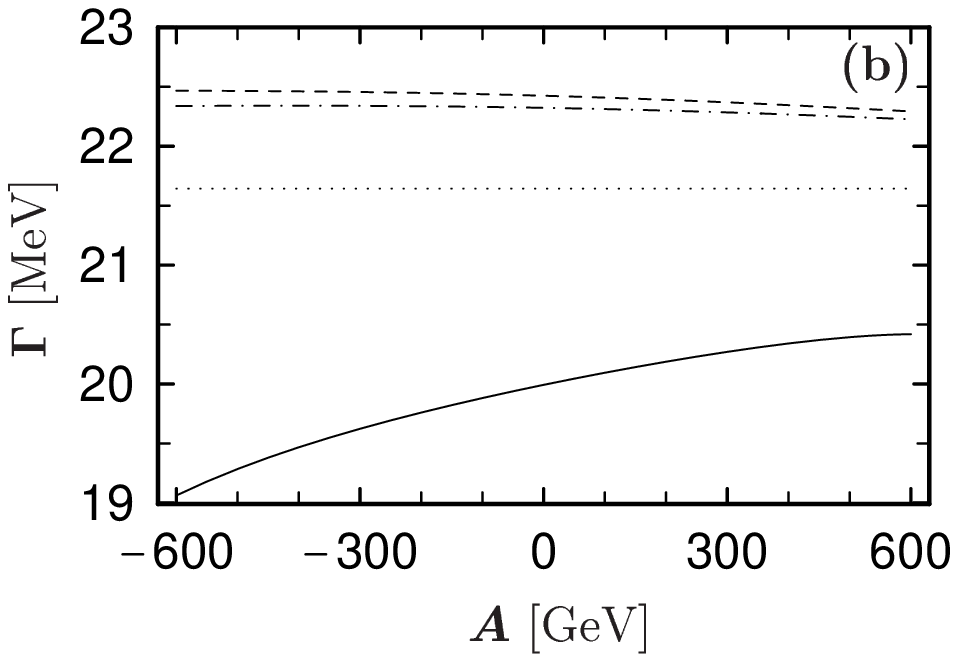}}}
 \hspace{-10mm}
 \vspace{-0mm}
 \caption[fig5]
{The widths of the decays $H^0\to\ch^0_1+\ch^0_2$~(a) and
$A^0\to\ch^0_1+\ch^0_3$~(b) as a function of $A$. The dotted line
corresponds to the tree-level width, the dash-dotted, dashed and
solid line corresponds to the ``conventional'', ``conventional +
$\Delta{Z}$'', and full correction, respectively. The parameters
are $\tan{\beta}=10$, $M=500$~GeV, and $\mu=150$~GeV~(a) and
$\tan{\beta}=50$ and $M=\mu=300$~GeV~(b).}
 \label{fig:Adependence}
 \end{center}
 \vspace{-5mm}
 \end{figure}

Figs.~\ref{fig:Adependence}a and \ref{fig:Adependence}b show the
dependence of the widths on the trilinear coupling $A$ for the
same decays modes and parameter sets as in
Figs.~\ref{fig:MsQdependence}a and \ref{fig:MsQdependence}b,
respectively. The $A$ dependence is mainly caused by $A_t$ and numerically important
in general.

\begin{figure}[h!]
 \begin{center}
 % \vspace*{-10mm}
 \hspace{-8mm}
 \mbox{\resizebox{84mm}{!}{\includegraphics{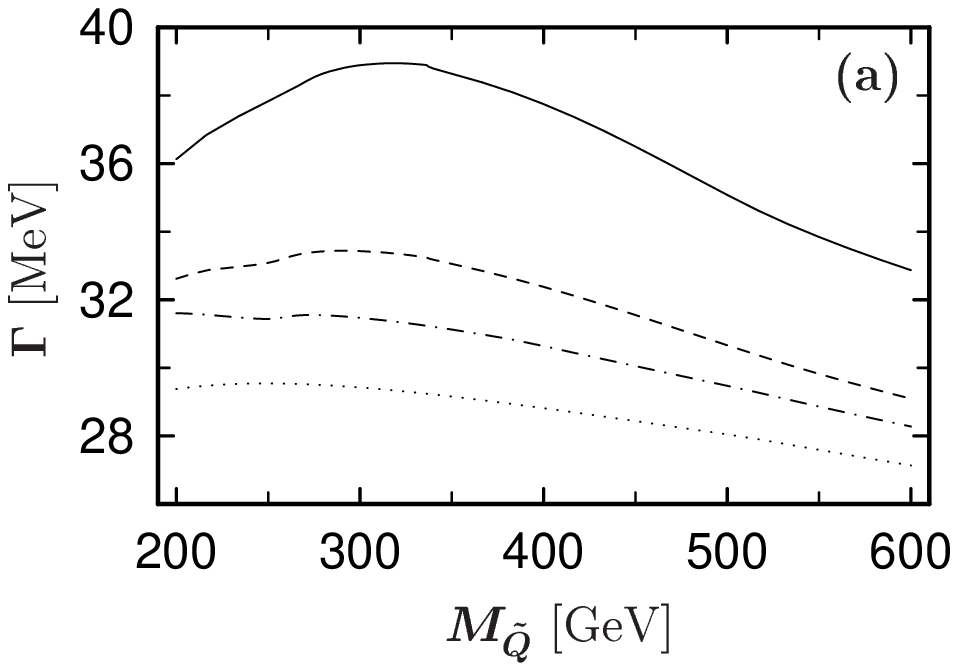}}}
 \hspace{-4mm}
 \mbox{\resizebox{85mm}{!}{\includegraphics{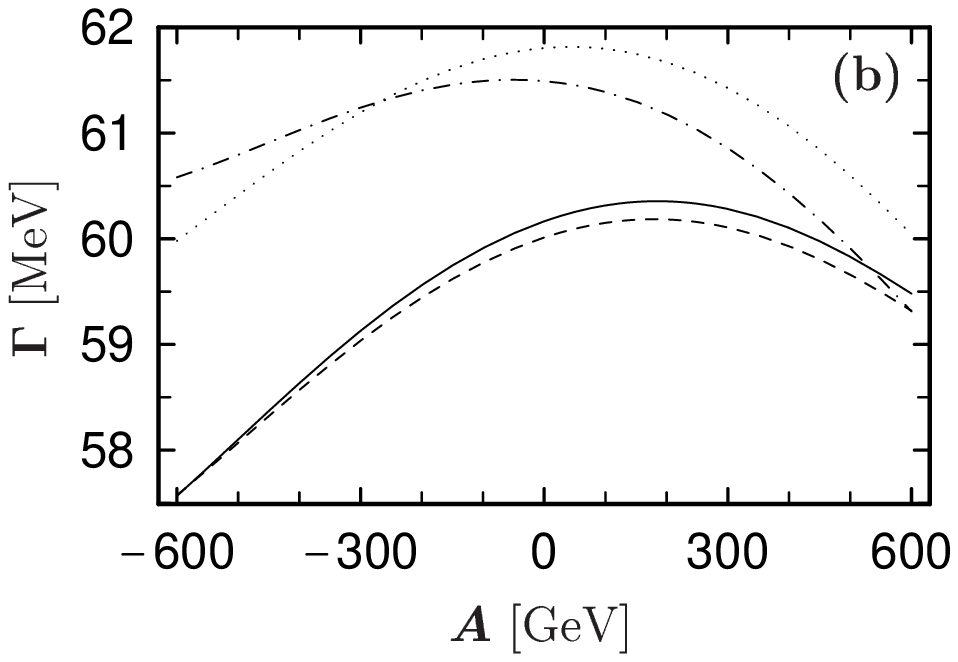}}}
 \hspace{-10mm}
 \vspace{-0mm}
 \caption[fig5]
 {The widths of the decays $\ch^0_3\to h^0+\ch^0_1$~(a) and
 $\ch^0_4\to h^0+\ch^0_1$~(b) as a function of $M_{\tilde{Q}}$~(a) and
 $A$~(b).
 The dotted line corresponds to the tree-level width, the dash-dotted,
 dashed and solid line correspond to the ``conventional'',
 ``conventional + $\Delta{Z}$'', and full corrections, respectively.
 The parameters are $\tan{\beta}=10$, $M=500$~GeV, and $\mu=150$~GeV~(a)
 and $\tan{\beta}=10$ and $M=\mu=300$~GeV~(b). }
 \label{fig:NeuDecay}
 \end{center}
 \vspace{-5mm}
 \end{figure}

We also discuss the related decays (\ref{eq:chidecay}) of the neutralinos.
In Figs.~\ref{fig:NeuDecay}a and \ref{fig:NeuDecay}b we show the
corrections to the width of the decays $\ch^0_3\to h^0+\ch^0_1$
and $\ch^0_4\to h^0+\ch^0_1$ as functions of $M_{\tilde{Q}}$ and
$A$, respectively. The parameters are as in
Fig.~\ref{fig:MsQdependence}a for Fig.~\ref{fig:NeuDecay}a and
$\tan{\beta}=10$ and $M=\mu=300$~GeV for Fig.~\ref{fig:NeuDecay}b.
The total correction can go up to 25\%.

\begin{figure}[h!]
 \begin{center}% \vspace*{-10mm}
 \hspace{-12mm}
 \mbox{\resizebox{78mm}{!}{\includegraphics{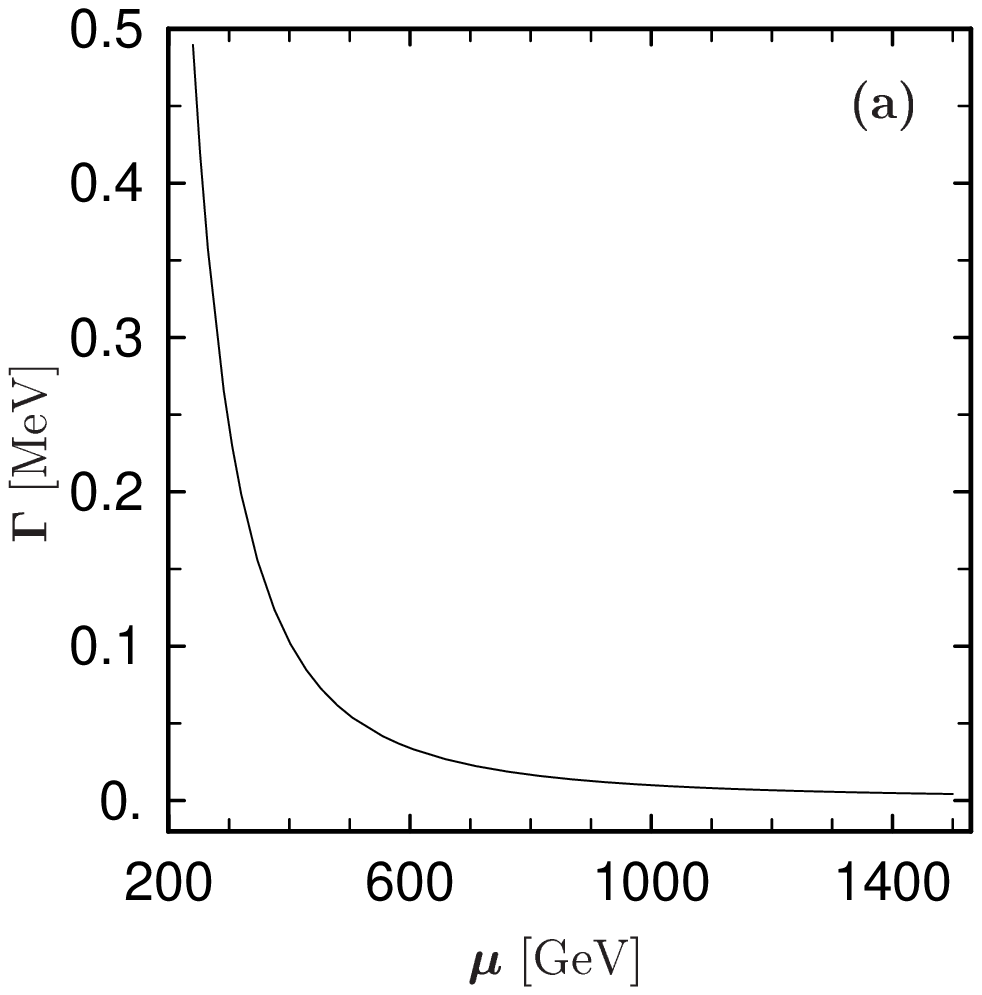}}}
 \hspace{-0mm}
 \mbox{\resizebox{78mm}{!}{\includegraphics{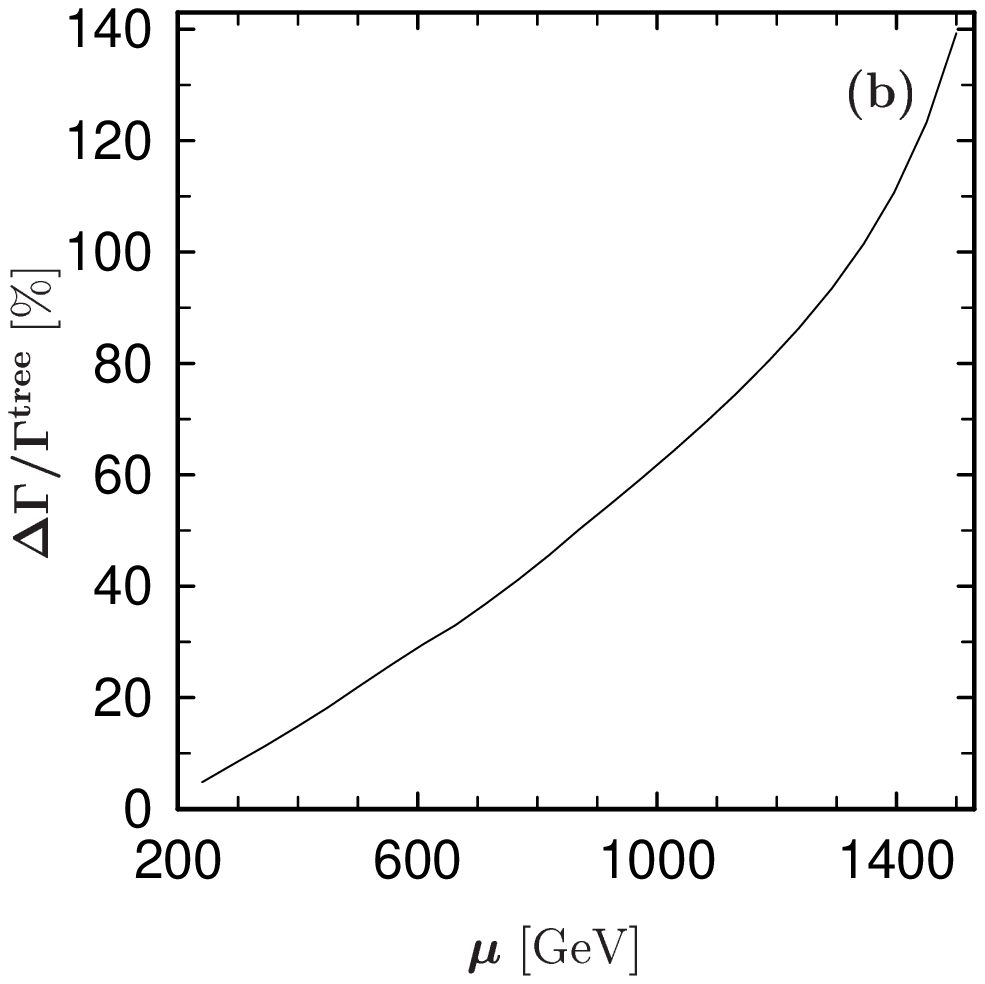}}}
 \hspace{-10mm}
 \vspace{-0mm}
 \caption[fig6]
 {The tree-level width (a) and the relative full correction~(b) of the
 decay\linebreak
 $h^0\to\ch^0_1+\ch^0_1$ as a function of $\mu$ for the parameters $\tb=20$,
 $\{M,\, M',\, m_{\tilde{g}},\, m_{A^0}\}=\{120,\,30,\,500,\,300\}$~GeV,
 $\{A_t,\,A_b,\,A_{\tau}\}=\{1,\,0,\,0\}$~TeV, and
 $M_{\tilde Q_{1,2}}=M_{\tilde U_{1,2}}=M_{\tilde D_{1,2}}=
 M_{\tilde Q_{3}}=M_{\tilde U_{3}}=M_{\tilde D_{3}}=
 2M_{\tilde L_{1,2}}=2M_{\tilde E_{1,2}}=
 2M_{\tilde L_{3}}=2M_{\tilde E_{3}}=500$~GeV. }
 \label{fig:lighthiggsdecay}
 \end{center}
 \vspace{5mm}
 \end{figure}

Finally, Fig.~\ref{fig:lighthiggsdecay} shows the $\mu$ dependence
of the width of the decay $h^0\to\ch^0_1+\ch^0_1$
(\ref{eq:h0ch1ch1}), both the tree-level value
(Fig.~\ref{fig:lighthiggsdecay}a) and the relative one-loop full
correction (Fig.~\ref{fig:lighthiggsdecay}b). This decay occurs
when $\tilde{\chi}^0_1$ is sufficiently light and is mainly a U(1)
gaugino to escape from the present direct search. In order to
realize this case, we consider very small $M'$ and take the
following parameters which are similar to those in
Ref.~\cite{Djouadi-Drees}:
$M_{\tilde Q_{1,2}}=M_{\tilde U_{1,2}}=M_{\tilde D_{1,2}}=
M_{\tilde Q_{3}}=M_{\tilde U_{3}}=M_{\tilde D_{3}}=
2M_{\tilde L_{1,2}}=2M_{\tilde E_{1,2}}=
2M_{\tilde L_{3}}=2M_{\tilde E_{3}}=500$~GeV,
$A_t=1000$~GeV, $A_b=A_{\tau}=0$~GeV, $M'=30$~GeV, $M=120$~GeV,
$m_{\tilde{g}}=500$~GeV, $\tb=20$, and $m_{A^0}=300$~GeV. The loop
correction can be comparable to or even larger than the tree-level
width as observed in Ref.~\cite{Djouadi-Drees}. Although the decay
width is much smaller than the other modes, the effect of the loop
correction might be seen in precision studies of $h^0$ at a linear
collider~\cite{Djouadi-Drees} and in neutralino dark matter
search~\cite{Djouadi-Drees,dark1,dark2}.

\section{Conclusions }
\label{sec:concl}

We have presented the calculation of the one-loop corrections to
the decays\linebreak $(h^0, H^0, A^0) \to \ch^0_m  + \ch^0_n$ and
$\ch^0_m \to (h^0, H^0, A^0)  + \ch^0_n$, $(m,n=1,\ldots,4)$, with
all fermions and sfermions in the loop. These decays are special
in the sense that they require particular care in the treatment of
the neutralino mixing and mass matrix in a scheme, where all
parameters in the neutralino mass matrix $Y$ and mixing matrix $Z$
are defined on-shell. We have shown the importance of the
corrections to these matrices in addition to the conventional
corrections (vertex and wave-function corrections with counter
terms). We have studied the dependence on the parameters $M$,
$\mu$, $A$, $M_{\tilde Q}$, and $\tan\beta$. The corrections to
the widths of the decays $(h^0, H^0, A^0) \to \ch^0_m  + \ch^0_n$
can go up to $\sim$ 15\%, those of the decays $\ch^0_m \to (h^0,
H^0, A^0)  + \ch^0_n$ to $\sim$ 25\%. For the invisible decay $h^0
\to  \ch^0_1  + \ch^0_1$, giving up the GUT relation for $M'$,
one even gets corrections up to 140\%.

\section*{Acknowledgements}

The work of Y.\,Y. was supported in part by the Grant--in--aid for
Scientific Research from Japan Society for the Promotion of
Science, No.~12740131. The work was also supported by the ``Fonds
zur F\"orderung der wissenschaftlichen Forschung'' of Austria,
project no.~P13139-PHY and the EU TMR Network Contract
HPRN-CT-2000-00149.

\begin{appendix}
\section*{Appendix}
\label{sec:app} \setcounter{equation}{0}
\renewcommand{\theequation}{A.\arabic{equation}}

In the following we give the formulae for the couplings and for
$\d e$. Furthermore, the proof of the gauge independence of the
processes considered will be given.

\section{Coupling parameters\label{sec:appA}}

The $H^0_k\,\cch_m\,\cch_n$ interaction is given by
\begin{equation}
  {\cal L} = - \frac12\, g\, H^0_a\,\bar\ch^0_m\,F^0_{mna}\,\cch_n +
  \frac{i}{2}\, g\, H^0_c\,\bar\ch^0_m\,F^0_{mnc}\,\gamma^5\,\cch_n\, ,
\label{eq:F0lag}
\end{equation}
with $a = 1,2$, $c = 3,4$, and
\begin{eqnarray}
 F^0_{mnk}&=& \hphantom{+} \frac{d^u_k}{2} \,
   \bigg[Z_{m4}Z_{n2}+Z_{n4}Z_{m2}-
 \tw(Z_{m4}Z_{n1}+Z_{n4}Z_{m1})\bigg]
 \nonumber\\&&
 + \frac{d^d_k}{2} \, \bigg[Z_{m3}Z_{n2}+Z_{n3}Z_{m2}-
 \tw(Z_{m3}Z_{n1}+Z_{n3}Z_{m1})\bigg]\,.
 \label{eq:F0tree}
\end{eqnarray}
The $Z_{mn}$ are the elements of the neutralino mixing matrix
which diagonalizes the neutralino mass matrix and
 \begin{eqnarray}
 d_k^4 = d^u_k&=&
 \Big(-\ca,\,-\sa,\,\hphantom{-}\cbe,\,\hphantom{-}\sbe\Big)_k\,,
 \nonumber\\
 d_k^3 = d^d_k&=&
 \Big(-\sa,\,\hphantom{-}\ca,
 \,-\sbe,\,\hphantom{-}\cbe\Big)_k\,.
 \label{eq:dudd}
 \end{eqnarray}
The superscript ``$u$'' denotes an up-type and ``$d$'' a down-type fermion.

The neutral Higgs boson-fermion-fermion couplings, defined by ${\cal L} = s^f_k\,
H_k^0\, \bar f \, f$ are
\begin{equation}
  s^{u}_a = \frac{h_{u}}{\sqrt{2}}\;d^{u}_a\,,
 \hspace{10mm}
 s^{u}_c = i\,\frac{h_{u}}{\sqrt{2}}\;d^{u}_c\,,
  \hspace{10mm}
  s^{d}_a = -\,\frac{h_{d}}{\sqrt{2}}\;d^{d}_a\,,
  \hspace{10mm}
 s^{d}_c\;=\;-\,i\,\frac{h_{d}}{\sqrt{2}}\;d^{d}_c\,,
 \label{eq:sfk}
\end{equation}
with $a = 1,2$ and $c = 3,4$, using the Yukawa couplings
\begin{equation}
  h_u = \frac{g\, m_u}{\sqrt2\, m_W \sin\beta}\,,
  \hspace{10mm}
  h_d = \frac{g\, m_d}{\sqrt2\, m_W \cos\beta}\, .
  \label{eq:yuk}
\end{equation}

The fermion-sfermion-neutralino coupling parameters $a^{\sf}_{ip}$ and
$b^{\sf}_{ip}$ ($i = 1,2,\, p = 1,\ldots,4$) have the form
\begin{eqnarray}
  \label{eq:neufsfcoup}
a^{\sf}_{ip} &=& g f_{Lp}^f R^\sf_{i1} +
 h_f Z_{px} R^\sf_{i2}\,,
 \\
  \label{eq:neufsfcoup1}
b^{\sf}_{ip} &=&
 g f_{Rp}^f R^\sf_{i2} +
 h_f Z_{px}R^\sf_{i1}\,,
\end{eqnarray}
with $x=3$ for down-type and $x=4$ for up-type fermions, $R^\sf$
the $2 \times 2$~sfermion rotation matrix,
 %and $Z$ the $4 \times 4$~neutralino rotation matrix,
\begin{eqnarray}
  \label{eq:fLi}
f_{Lp}^f &=& \sqrt2\, \bigg[ \Big(e_f-\IL\Big) \tw Z_{p1}+\IL Z_{p2}\bigg]\,,
 \\
 \label{eq:fRi}
f_{Rp}^f &=& -\sqrt2\, e_f \,\tw\, Z_{p1}\,.
\end{eqnarray}
$\IL$ denotes the SU(2)$_L$ isospin and $e_f$ the charge of the fermion $f$.

The $H^0_k\, \sf_i^*\, \sf_j$ couplings~\cite{gunion} are
\begin{eqnarray}
\label{eq:7}
 && \hspace*{-2cm}
 G_{ij1}^\su \equiv G(h^0\, \su_i^*\, \su_j) =
\left(R^\su\, G^\su_{LR1}\, (R^{\su})^T \right)_{ij} = \nonumber\\
 && \hspace*{-2cm}
 -g R^{\su} \left(
\begin{array}{cc} -\frac{m_Z}{c_W}(I_u^{3 L} - e_u s_W^2) \si_{\al + \be} +
\frac{m_u^2}{m_W\si_\be} \co_\al & \frac{m_u}{2 m_W \si_\be} (A_u \co_\al + \mu
\si_\al)\\ \frac{m_u}{2 m_W \si_\be} (A_u \co_\al + \mu \si_\al) &
-\frac{m_Z}{c_W} e_u s_W^2 \si_{\al + \be} + \frac{m_u^2}{m_W \si_\be} \co_\al
\end{array}
\right)(R^{\su})^T\, ,
\end{eqnarray}
\begin{eqnarray}
 \label{eq:8}
 && \hspace*{-2.9cm}
 G_{ij1}^\sd \equiv G(h^0\, \sd_i^*\, \sd_j) = \left(R^\sd\, G^\sd_{LR1} \,
(R^{\sd})^T \right)_{ij} =  \nonumber\\
 && \hspace*{-2.9cm}
 g R^{\sd}
\left( \begin{array}{cc} \frac{m_Z}{c_W}(I_d^{3 L} - e_d s_W^2) \si_{\al + \be}
+ \frac{m_d^2} {m_W\co_\be} \si_\al & \frac{m_d}{2 m_W \co_\be} (A_d \si_\al +
\mu \co_\al)\\ \frac{m_d}{2 m_W \co_\be} (A_d \si_\al + \mu \co_\al) &
\frac{m_Z}{c_W} e_d s_W^2 \si_{\al + \be} + \frac{m_d^2}{m_W \co_\be} \si_\al
\end{array}
\right)(R^{\sd})^T\, ,
\end{eqnarray}
\begin{eqnarray}
 & G_{ij2}^\sf \equiv G(H^0\, \sf_i^*\, \sf_j) =  G(h^0 \sf_i^* \sf_j)
\mbox{ with } \al \to \al - \frac\pi2\,,
 \label{eq:9}
\end{eqnarray}
(i. e. $\ \sin\al \equiv \si_\al \to -\co_\al\,,\ \cos\al \equiv \co_\al \to
\si_\al\,,\ \mbox{and}\; \sin(\al+\be) \equiv \si_{\al+\be} \to
-\co_{\al+\be})$,

\begin{eqnarray}
\label{eq:10}
 G_{ij3}^\su &=& G(A^0\, \su_i^*\, \su_j) = \frac{i g}{2 m_W}
\left(
\begin{array}{cc} \hphantom{-}0 & 1\\ -1 & 0
\end{array}
\right)_{ij}\, m_u (A_u \cot\be + \mu)\, ,
\end{eqnarray}
\begin{eqnarray}
 \label{eq:11}
  G_{ij3}^\sd &=& G(A^0 \sd_i^* \sd_j) = \frac{i g}{2 m_W} \left(
\begin{array}{cc}  \hphantom{-}0 & 1\\ -1 & 0
\end{array}
\right)_{ij}\,m_d (A_d \tan\be + \mu)\, ,
\end{eqnarray}
\begin{eqnarray}
 & G_{ij4}^\sf \equiv G(G^0\, \sf_i^*\, \sf_j) =  G(A^0 \sf_i^* \sf_j)
\mbox{ with } \be \to \be - \frac\pi2\,,
 \label{eq:12}
\end{eqnarray}
(i. e. $\ \tan\be \leftrightarrow -\cot\be $). The
superscript ``$\su$'' (``$\sd$'') denotes an up-type (down-type) sfermion.

The $H^0_k\,H^0_l\,\sf_i^*\,\sf_j$ interaction is given by
\begin{equation}
  {\cal L} = - \frac12\,\left( h_f^2\, c_{kl}^\sf\, \d_{ij} + g^2\,d_{kl}\, e_{ij}^\sf\right)\,
H^0_k\,H^0_l\,\sf_i^*\,\sf_j\, ,
\end{equation}
with
\begin{eqnarray}
 c^\su_{kl}&=&
 \left(\begin{array}{cccc}
 \cas & \hphantom{-}\ca\,\sa & 0 & 0\\
 \hphantom{-}\ca\,\sa & \sas & 0 & 0 \\
 0 & 0 & \cbs & \hphantom{-}\cbe\,\sbe \\
 0 & 0 & \hphantom{-}\cbe\,\sbe & \sbs  \\
 \end{array}\right)_{kl}\,,\\[3mm]
 c^\sd_{kl}&=&
 \left(\begin{array}{cccc}
 \sas & -\ca\,\sa & 0 & 0\\
 -\ca\,\sa & \cas & 0 & 0 \\
 0 & 0 & \sbs & -\cbe\,\sbe \\
 0 & 0 & -\cbe\,\sbe & \cbs  \\
 \end{array}\right)_{kl}\,,\\[3mm]
  d_{kl} & = & c^\sd_{kl} - c^\su_{kl}\,,\\[3mm]
  e_{ij}^\sf & = & \frac{1}{2\, c_W^2}\,\left[(I^{3L}_f - e_f
  s_W^2)\, R_{i1}^\sf\, R_{j1}^\sf +
e_f\, s_W^2\, R_{i2}^\sf\, R_{j2}^\sf \right]\, .
\end{eqnarray}

\setcounter{equation}{0}
\renewcommand{\theequation}{B.\arabic{equation}}
\section{Counter term \boldmath $\d e$\label{sec:appB}}
When we give the renormalized electric charge in the Thomson limit
with the measured fine structure constant $\alpha=e_0^2/(4\pi)=1/137$,
the counter term $\delta e_0$ is given by the general form \cite{denner}
\begin{equation}
\frac{\d e_0}{e_0} =\frac12\;\dot{\Pi}^{AA}_T(0)-\frac{s_W}{c_W}\,
\frac{\Pi^{AZ}_T(0)}{m_Z^2}\, , \label{eq:dZe}
\end{equation}
with the momentum derivative of the transverse photon self-energy
$\dot{\Pi}^{AA}_T$ and the $\gamma-Z^0$ mixing self-energy, both
for the on-shell photon ($p^2=0$). Fermions and sfermions do not
contribute to $\Pi^{AZ}_T(0)$ as a consequence of the fact that
the physical photon is massless to all orders. However, the
contribution of light hadrons to $\dot{\Pi}^{AA}_T(0)$ has a large
theoretical uncertainty
\cite{alphamz1,denner}.
To avoid this problem, in this work we use the \msbar running
coupling at $Q=m_Z$, $\alpha(m_Z)=e^2(m_Z)/(4\pi)=1/129$ as input.
The counter term $\delta e(m_Z)$ then becomes
    \begin{equation}
  \frac{\d e}{e} = \frac{e^2\, e_f^2}{(4\pi)^2}\, \left[
  \frac{2}{3}\left(
  \Delta + \log\frac{Q^2}{x_f^2} \right) +
 \sum_{i=1}^2\frac{1}{6}\left(
  \Delta + \log\frac{Q^2}{m_{\sf_i}^2} \right)\right] \, ,
 \label{eq:de/e2}
\end{equation}
with $x_f=m_Z$ for all $m_f<m_Z$ and $x_t=m_t$. Here $\Delta$
denotes the UV divergence factor.

\setcounter{equation}{0}
\renewcommand{\theequation}{C.\arabic{equation}}
\section{Proof of the \boldmath $\xi$ independence}
We investigate the wave-function corrections to the process
\begin{equation}
  A^0(p) \to \ch^0_m (k_1) + \ch^0_n (k_2) \, .
\label{eq:A0neu}
\end{equation}
Both the contributions of the transitions $A^0 \to G^0 \to \ch^0_m + \ch^0_n$
and $A^0 \to Z^0 \to \ch^0_m + \ch^0_n$ have a dependence on the gauge
parameter $\xi= \xi_Z$ in the propagators of $(G^0,Z^0)$.
We show that the sum of these contributions is independent of $\xi$. \\
We start from the matrix elements in a general $R_{\xi}$ gauge,
\begin{eqnarray}
  \M^G & = & ( i\, \Pi_{AG}) \frac{i}{p^2 - \xi m_Z^2} (- g F_{mn4}^0 )
  \bar u(k_1) \gamma^5 v(k_2)\, ,\\
\label{eq:MG1}
   \M^Z &  = & ( -i p^\mu \Pi_{AZ}) \frac{i}{p^2 - m_Z^2}
  \left( - g_{\mu\nu} + (1 - \xi)
  \frac{p_\mu p_\nu}{p^2 - \xi m_Z^2}\right) (-i \frac{g}{c_W} O^{''L}_{mn})
  \bar u(k_1) \gamma^\nu \gamma^5 v(k_2)   \, .\nonumber\\
\label{eq:MZ1}
\end{eqnarray}
The self-energies $\Pi_{AG}$ and $\Pi_{AZ}$ by (s)fermion one-loop
contributions are $\xi$ independent.
The $Z^0\ch^0_m\ch^0_n$ couplings $O^{''L}_{mn}$ are
\begin{equation}
\hspace*{-1cm} O^{''L}_{mn} = \frac{1}{2} (-Z_{m3} Z_{n3} + Z_{m4} Z_{n4})\, .
   \label{eq:Oppdef}
\end{equation}
As limiting cases, $\M^{G} = 0$ in the physical unitary gauge
$\xi \to \infty$ and $\M^{Z} = 0$ in the $\xi = 0$ (Landau) gauge.
Note that the tadpole contributions have to be
included \cite{pokorski,dabelstein} in $\Pi_{AG}$.

We can write $\M^G$ directly as
\begin{equation}
    \M^G  =  \frac{1}{p^2 - \xi m_Z^2}\,g\, \Pi_{AG}(p^2)\, F_{mn4}^0\,
  \bar u(k_1) \gamma^5 v(k_2)\, .
\label{eq:MG2}
\end{equation}
For $\M^Z$ we first contract the Lorentz indices,
\begin{eqnarray}
 &&\hspace*{-1cm}  p^\mu \,\left( - g_{\mu\nu} + (1 - \xi)
  \frac{p_\mu p_\nu}{p^2 - \xi m_Z^2}\right)
  \bar u(k_1) \gamma^\nu \gamma^5 v(k_2)
= \left( \frac{(1 - \xi)\, p^2}{p^2 - \xi m_Z^2} - 1\right)\, \bar u(k_1)\psla
\gamma^5 v(k_2) \, ,\nonumber
 \label{eq:tmp1}
\end{eqnarray}
and use $\bar u(k_1)\psla \gamma^5 v(k_2) = \bar u(k_1) (\ksla_1 +
\ksla_2) \gamma^5 v(k_2) = (m_m + m_n) \bar u(k_1)  \gamma^5
v(k_2)$. So we get
\begin{equation}
  \M^Z = - \frac{i}{p^2 - m_Z^2}
  \left( \frac{(1 - \xi)\, p^2}{p^2 - \xi m_Z^2} - 1\right)
  \frac{g}{c_W} \, \Pi_{AZ}(p^2)\, O^{''L}_{mn}\,
   (m_m + m_n)\,
   \bar u(k_1)  \gamma^5 v(k_2)\, . \label{eq:MZ2}
\end{equation}
We use the Slavnov-Taylor identity (see also \cite{dabelstein}, eq.~(3.7))
\begin{equation}
  p^2\, \Pi_{AZ}(p^2) + i\, m_Z\,  \Pi_{AG}(p^2) = 0\,,
\label{eq:wardID}
\end{equation}
and split the sum $\M^G + \M^Z$ in an obviously $\xi$ independent and possibly
dependent part,
\begin{eqnarray}
\hspace*{-0.8cm}  \M^{G + Z}
   & = &  \bigg\{ \frac{i}{p^2 - m_Z^2}\frac{g}{c_W} \,
  \Pi_{AZ}(p^2)\, O^{''L}_{mn}\,
  (m_m + m_n) \nonumber\\
  &&\hspace*{-2.2cm}+\, i\,g \, \Pi_{AZ}(p^2) \frac{p^2}{p^2 - \xi m_Z^2}\left(
\frac{F_{mn4}^0}{m_Z} -  \frac{(1 - \xi)}{p^2 - m_Z^2} \frac{O^{''L}_{mn}}{c_W}
(m_m + m_n) \right)\bigg\} \, \bar u(k_1)  \gamma^5 v(k_2) \, .
  \label{eq:MGplusZ1}
\end{eqnarray}
With the relation (proved later)
\begin{equation}
  m_Z\, c_W\, F^0_{mn4} = -  O^{''L}_{mn}\, (m_m + m_n)\,,
\label{eq:F02Opp}
\end{equation}
we get for the part written in the brackets in eq.~(\ref{eq:MGplusZ1})
\begin{equation}
 - (m_m + m_n) \frac{O^{''L}_{mn}}{c_W} \left( \frac{1}{m_Z^2}
 + \frac{1 - \xi}{p^2 - m_Z^2}\right) =
 - (m_m + m_n) \frac{O^{''L}_{mn}}{m_Z^2\,c_W}
 \frac{p^2 - \xi m_Z^2}{p^2 - m_Z^2}\, ,
\label{eq:tmp2}
\end{equation}
and therefore the final result
\begin{equation}
  \M^{G + Z} =- i\, \frac{g}{m_Z^2\,c_W}\,\Pi_{AZ}(p^2)\, O^{''L}_{mn}\,
  (m_m + m_n) \, \bar u(k_1)  \gamma^5 v(k_2)\, .
\label{eq:res1}
\end{equation}
The $\xi$ dependence is completely cancelled in (\ref{eq:res1}).

Finally, we prove (\ref{eq:F02Opp}).
With the abbreviation $A_{ij} = (Z_{mi} Z_{nj} + Z_{mj} Z_{ni})$ and knowing
the entries of the neutralino tree-level mass matrix $Y$
(see e.~g.~eq.~(35) in \cite{chmasscorr}), one can write
$F^0_{mn4}$ as
\begin{equation}
2\,m_Z\,c_W\,F^0_{mn4} = Y_{13} A_{31} + Y_{23} A_{32}
- Y_{14} A_{41} - Y_{24} A_{42} \,.
 \label{eq:tmp6}
\end{equation}
Next we add and subtract the terms $Y_{33} A_{33} + Y_{43} A_{34}$ and $Y_{34}
A_{43} + Y_{44} A_{44}$. Exploiting the fact that $Y_{33} = Y_{44} = 0$ and
$Y_{34} =  Y_{43}$, we get
\begin{equation}
 2\,m_Z\,c_W\,F^0_{mn4} =  \sum_k \left( Y_{k3} A_{3k} - Y_{k4} A_{4k} \right)
 \,.
 \label{eq:tmp7}
\end{equation}
Writing the entries of $Y$ in terms of neutralino masses, $Y_{kj} = \sum_l m_l
Z_{lk} Z_{lj}$, and using $\sum_k Z_{ik} Z_{jk} = \delta_{ij}$ we get
\begin{eqnarray}
2\,m_Z\,c_W\,F^0_{mn4} & = & \sum_{k,l} m_l Z_{lk} \left[Z_{l3}\left(Z_{m3}
Z_{nk} + Z_{mk} Z_{n3}\right) - Z_{l4}\left(Z_{m4} Z_{nk} + Z_{mk}
Z_{n4}\right)\right]
 \nonumber\\ & = &  \sum_{l} m_l \left( \delta_{nl} Z_{l3} Z_{m3} +
 \delta_{ml} Z_{l3} Z_{n3} - \delta_{nl} Z_{l4} Z_{m4} -
 \delta_{ml} Z_{l4} Z_{n4} \right) \nonumber\\ & = &
 m_n Z_{n3}Z_{m3} +  m_m Z_{m3}Z_{n3} - m_n Z_{n4}Z_{m4} -  m_m Z_{m4}Z_{n4}
 \nonumber \\
 & = & \left(Z_{m3} Z_{n3} - Z_{m4}Z_{n4}\right) (m_m + m_n) =
-  2\, O^{''L}_{mn}\, (m_m + m_n)\,.\label{eq:tmp8}
\end{eqnarray}
Therefore, eq.~(\ref{eq:F02Opp}) is proven.

However, from the Slavnov-Taylor identity one can prove in general that the
same cancellation of the gauge dependent parts in $G^0$ and $Z^0$
propagators occurs for any one-loop two-body decay of $A^0$.

\end{appendix}

\end{document}